# The Venus Hypothesis

**Author:** A. Cartwright [1]

**Affiliations:**

[1]School of Physics and Astronomy, Cardiff University

*Correspondence to: cartwrighta@cf.ac.uk

**Abstract**: Current models indicate that Venus may have been habitable. Complex life may have evolved on the highly irradiated Venus, and transferred to Earth on asteroids. This model fits the pattern of pulses of highly developed life appearing, diversifying and going extinct with astonishing rapidity through the Cambrian and Ordovician periods, and also explains the extraordinary genetic variety which appeared over this period.

**One Sentence Summary:** Unanswered questions arising from the sudden appearance and radiation of complex life during the Cambrian and Ordovician periods, are answered by the proposal that these complex life forms evolved on Venus, and were transferred here by lithopanspermia.

**Habitability of Venus and transport of material from Venus to Earth**

The theory of lithopanspermia was introduced in the 1980s(*1*) and is now widely accepted. The head of NASA's search for life on Mars and Europa commented this year "There has almost certainly been exchange of biota between Earth and Mars. Mars is not very far away. Martian rocks from impact can easily reach Earth; probably Earth rocks also reached Mars from impacts in the past" (*2*). The routes by which life can have been transported between Earth, Mars and the moons of Jupiter have been modelled(*3*), and plants, bacteria and primitive animals such as tardigrades have been shown to survive the vacuum of space (*4-6*). The possibility that Venus might have been the source of life, which then travelled to Earth, seems not to be considered by those working in the field, probably because Venus is currently so inhospitable. It may not always have been so, and it is closer than Mars.

Current 3d models of the atmosphere of Earth-sized planets orbiting closer to their stars have shown that the habitable zone extends closer to the stars than previously thought. The most recent research brings forward the habitability of Venus to at least 750My BP(*7*) and the authors speculate that life could have evolved. Current models of the interaction of crust and atmosphere on Venus, acknowledge that a habitable Venus cannot be ruled out (*8*). Isotope studies(*9*) also indicate the previous existence of liquid water on Venus. If Venus did evolve to have liquid water, and then develop plate tectonics and a carbon cycle as has been suggested (*8*), then an Earth-like world could have evolved, with the carbon in the atmosphere being absorbed by weathering and forming carbonate rocks in the crust.

It is known that if plate tectonics was active early on a habitable Venus, then it must have ceased by the time the planet developed into the stagnant lid phase (*10*) with violent resurfacing occurring by about 600-400 My BP (*11-14*). These age estimates have an uncertainty factor of about 200%, as they are obtained from crater counting models(*14*). All of the Venusian crust is thought to have melted and then resolidified at that time. Any carbonate rocks existing before this happened would have been reprocessed in the mantle of Venus, releasing their $CO_2$ just as happens on Earth, but on an enormous scale, leading to a runaway greenhouse effect and the evaporation of water, with the concomitant negative implications for life.

The stagnant lid period would have seen increasing pressure beneath the crust(*10*) as heat transfer by conduction through the Venusian crust was no longer efficient enough to balance the radioactive heating within the planet's interior. Occasional volcanic eruptions and impacting asteroids would therefore have resulted in more explosive ejecta launch. This would be a perfect scenario to increase the amount of material being launched from Venus' crust to escape velocity and even to Earth-crossing orbit. The escape velocity for Venus is around 10km/s and an additional impulse of 22km/s would be required for an object to just reach Earth orbit (Hohman Transfer orbit calculation) with a transfer time of only 150 days.

Recent models (*8*) suggest that the resurfacing event would have taken approximately 100 My to complete, indicating that it was taking place between 700My and 400 My BP.

**Cambrian explosion and Great Ordovician Biodiversity Event.**

Complex life appeared suddenly in Earth's fossil record with the Cambrian Explosion(*15, 16*) of 540My BP, when all but one of the known phyla appeared on Earth. This was followed by the Great Ordovician Biodiversity Event (GOBE) in the period up to 460My BP, when diversity greatly increased(*17*). Land plants also appeared around 470My BP (*18*). The timing is suggestively close to the resurfacing event on Venus.

The nature and rapidity of the Cambrian explosion represent one of the major challenges in the study of evolution on Earth (*19*). Any model of the Cambrian explosion, and the GOBE, must explain the timing of this period – why did it start, and why did it stop? Why is there a lack of early fossil precursors of species which can be shown by genetic analysis to have evolved 1 billion years ago (*20*)? In particular, what could have caused all 34 known phyla to appear during the Cambrian or just shortly afterwards, with no new phyla ever evolving subsequently? Why was this period unique? The GOBE presents the additional challenge that timings of significant changes are 'diachronous across groups, environments and regions'(*17*).

A systematic review of all of these problems and possible solutions(*21*) showed that none of the current environmental, developmental or ecological theories convincingly answer all of the questions. Much work is currently being focused on ocean chemistry and the increase in oxygen levels (*22*), acting as a trigger to sudden development and the rise of predation, but this is not conclusive at present. The sporadic arrival of complex life from Venus, during the lead up to and early stages of its violent resurfacing, is suggested here to provide an elegant solution to these outstanding questions.

The timing of the arrival of material from Venus was dictated by the slowing and cessation of plate tectonics on that planet, leading to the stagnant lid state and eventual violent resurfacing(*8, 10*), possibly exacerbated by the break up of the L-chondrite parent asteroid(*23*). The lack of precursor fossils arises because the precursors were left behind on Venus. The geographical inconsistency of the GOBE, with some species suddenly flourishing in one area while elsewhere dying out, would be expected, as some meteorites might deliver new lifeforms, but others might have very destructive effects on established communities.

The fantastic genetic diversity, much of which died out never to evolve again, would be predicted to arise on Venus due to its proximity to the Sun. Venus is subjected to twice the intensity of solar radiation and solar wind experienced on Earth, while the tidal derotation has led to very long day lengths(*24*), and hence a negligible magnetic field. Without the protection of a magnetosphere, life on Venus would have been bombarded by the Solar wind. Exposure to radiation leads to accelerated mutation rates(*25*) . If life evolved on Venus, therefore, we would predict that it would be wildly more exotic and further advanced than that evolving on Earth. It would be plausible for Venusian and Earth life to share common ancestry, through earlier periods of lithopanspermia (*1, 2*).

Another key difference between Venus and Earth, leading to more genetic variety, would be the smaller Venusian water volume, estimated at only 10% of Earth's(*7*). If Venus therefore had more land and less ocean, then unlike Earth, with contiguous oceans and islands of land, Venus would be more likely to have contiguous land with large isolated bodies of water. We know from Earth's evolution that islands (New Zealand, Australia, Madagascar…) do not have to be isolated for very long before they develop very distinctive fauna. On Venus, this could have happened in the oceans.

The increasing day length on Venus, extending perhaps to years then decades, would give a very definite evolutionary advantage to life evolving extended, deep hibernation states, or diapauses, with high resistance to extreme temperatures and levels of radiation. Some of our most ancient life forms, such as tardigrades(*5*), nematodes(*6*) and triops canciformis(*26*), exhibit just those features, and would therefore be well equipped to survive the transfer to Earth. Life on the dark, night side of Venus, most likely to be hit by impactors and then launch material to Earth, would be deeply dormant, and probably encased in ice. Ideally prepared for interplanetary travel.

**Evidence of impacts**

An explosive impact in upper New York state at 540My BP has been detected, indicated by high Iridium levels in the relevant rock formations. This has been explicitly linked to the Cambrian explosion (of life) (*27*).

The Great Ordovician Biodiversity Event has been explicitly linked (*28*) to the break-up of the large L-chondrite asteroid at 470My BP(*23, 29*). Remnants of L chondrites embedded in limestone, as well as analysis of extra- terrestrial osmium and chromite levels, have demonstrated the precise coincidence between asteroid flux and biodiversification, in Scandinavia and China(*28*).

The rate of impacts during this period is believed to have increased by 1 to 2 orders of magnitude(*28*). 5 large impact craters in North America and 6 in Scandinavia have been dated to the Ordovician period(*30*), and these impactors would have been accompanied by many times their volume of smaller pieces of debris, many of which may have landed in shallow water and survived impact(*31*). Asteroids do not burn up during their passage through the atmosphere. The outer layers burn away or spallate, but the insides remain cold. Medium sized blocks could have landed in water, perhaps splitting open as they finally bumped into the seabed, as happened with the 950kg Chelyabinsk meteorite(*31*), and releasing dormant life forms into a habitable new environment.  These blocks would consist of rocks, including carbonates, from the crust of Venus.

As well as craters and the chemical evidence of extraterrestrial impacts, rocks of the Ordovician period display 'odd lithologies and strange fauna… in China, unusual mini-mounds interrupt the normal succession'(*28*). Many limestone boulders are present in Scandinavia(*32*), and there are large Breccia fields throughout this period, which have again been explicitly linked with impacts(*33, 34*). I suggest that these carbonate boulders, mounds and breccia fields are remnants of Venusian asteroids, forming 'strewn fields' related to the many impact craters formed during the Ordovician.

Note that the L-chondrite asteroid break up and subsequent migration of asteroid fragments to Earth-crossing orbits(*23*) would have led not only to bombardment of Earth, but also Venus. An asteroid fragment travelling on an elliptical orbit with perihelion at Venus orbit and aphelion in the asteroid belt would encounter Venus at a higher velocity than it would Earth. The extra kinetic energy of the Venusian impacts, exacerbated by the high pressure developed under the Venusian crust in the 'stagnant lid ' phase(*10*), would have had a more dramatic effect on Venus than Earth. Indeed, the L-chondrite break-up and suggested bombardment of Venus may even have contributed to or accelerated the final resurfacing of Venus. Venusian material ejected would be imprinted with traces of L-chondrite impactor material when it arrived at Earth. Chondrites, chromite and osmium signatures would still be found associated with some of the impact sites of Venusian material.

**Implications for the search for life.**

I have suggested that life evolved simultaneously on Earth and Venus. On Earth, it evolved very slowly while on the neighbouring, more highly irradiated planet, complexity and variety arose much earlier. That planet became uninhabitable, but, during that process, through violent volcanic eruptions and asteroid impacts, living complex lifeforms transferred to Earth. Here, with our protective magnetosphere, ozone layer and steady climate, life evolved at a more stately pace, but emerged from the sea and developed intelligence. This is an elegant and simple solution to the many complex questions arising from the Cambrian Explosion and GOBE, satisfying the demands of Occam's razor. It does, however, further complicate the search for life in other solar systems, and reduce the odds, if a two stage evolution model is adopted.

**Acknowledgments:** I have been fortunate to be able to access the huge body of work by Birger Schmitz, not all of which can be cited here.